# On the Footsteps to Generalized Tower of Hanoi Strategy

Bijoy Rahman Arif

IBAIS University, Dhaka, Bangladesh

Email: bijoyarif71@yahoo.com

**Abstract**

In this paper, our aim is to prove that our recursive algorithm to solve the "Reve's puzzle" (four- peg Tower of Hanoi) is the optimal solution according to minimum number of moves. Here we used Frame's five step algorithm to solve the "Reve's puzzle", and proved its optimality analyzing all possible strategies to solve the problem. Minimum number of moves is important because no one ever proved that the "presumed optimal" solution, the Frame-Stewart algorithm, always gives the minimum number of moves. The basis of our proof is Bifurcation Theorem. In fact, we can solve generalized "Tower of Hanoi" puzzle for any pegs (three or more pegs) using Bifurcation Theorem. But our scope is limited to the "Reve's puzzle" in this literature, but lately, we would discuss how we can reach our final destination, the Generalized Tower of Hanoi Strategy. Another important point is that we have used only induction method to prove all the results throughout this literature. Moreover, some simple theorems and lemmas are derived through logical perspective or consequence of induction method. Lastly, we will try to answer about uniqueness of solution of this famous puzzle.

**Introduction**

The original "Tower of Hanoi" puzzle of three pegs was created by French mathematician Edouard Lucas in 1883. But we have to remember that similar problems in more or less similar forms were considered by many natural worshipers linked about their spiritual issues. Normally we are given a tower of eight disks, initially stacked in decreasing size on one of three pegs:

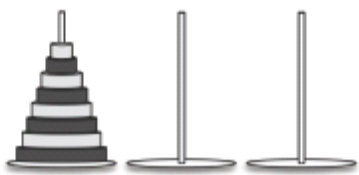

Fig. Illustration of the "Tower of Hanoi" *

The objective is to transfer the entire tower to one of the other pegs, moving only one disk at a time, and never moving a larger one onto a smaller. The minimum number of moves for a tower of $n$ disks was quickly shown to be $2^n-1$, with simple recursive solution [1, 6].

In 1908, Henry Ernest Dudeney published a book of puzzles which included a variation of standard Tower of Hanoi [2]. Dudeney used wheels of cheese instead of disks, but more importantly added a fourth peg, and named it as "The Reve's Puzzle". Here all pegs are identical and we have named them as Source, Mediator, Reservoir, and Destination respectively. We have to remember that all four pegs' role is

*Courtesy: Michael Rand

interchangeable, and there is no need to stick to any peg for particular task. But before a number of moves, we must ensure the role of all four pegs.

Then, in 1939 the American Mathematical Monthly published a generalized Tower of Hanoi problem, asking for a solution for *n* disks of any number *k* of pegs [3]. After two years, Frame and Stewart offered algorithms for solving the problem in a minimum number of moves, but an editorial note pointed out that neither had proved their algorithm correct [4]. Stewart's *3*-step algorithm is much simpler than Frame's *5*-step algorithm, but we will work with Frame's one, and will prove its optimality. However, the results of these two algorithms were later proven to be equivalent [5].

**Existence of Solution:**

Before starting solution of any problem, at first, we have to find out existence of solution of the problem. So we must show the generalized "Tower of Hanoi" puzzle is soluble for any number of pegs. At first, we prove the original "Tower of Hanoi" puzzle of three pegs can be solved for any number of disks. Then it follows that generalized "Tower of Hanoi" is soluble completely.

If we consider the original "Tower of Hanoi" with number of disks, *n = 1, 2, 3, ...* then minimum number of moves required *1, 3, 7, ...* respectively. Let for *n = m* disks, the tower is transferable. Now we consider a tower of *n = m+1* disks and show that it is also transferable.

As for *n = m* disks, the tower is transferable, for *n = m+1* disks, we can transfer first m disks to any other peg than Source, leaving behind the last biggest disk. Now Source has the last biggest disk only, one peg is empty, and last peg is equipped with first m disks sequentially. Hence we transfer the biggest disk to empty peg and transfer the first m disks over the biggest disk. It is possible because all pegs are identical as per rule of puzzle. So the original "tower of Hanoi" is soluble for any number of disks.

So we can conclude that the "Reve's Puzzle" or four pegs "Tower of Hanoi" is soluble for any number of disks. Eventually, it ensures the existence of solution of the generalized "Tower of Hanoi" puzzle.

**Blank Peg Theorem**

Before moving any disk (which has not been moved yet) from source peg following the puzzle's rules, we need at least a peg to be blank beside the Source peg.

*Proof.* The disk which we want to move last from Source peg is bigger than any other disks which are already moved. So we can't place the last disk on top of any other disks. Hence a peg must be blank beside the Source peg.  □

**Block-Unit Theorem:**

If we move any number of disks (not necessarily sequential) from any peg to other peg following the puzzle's rule, number of moves required to move the block of disks as a whole is the summation of moves required per unit disk as isolated.

*Proof.* According to puzzle constraints, we have to move a disk at a time. As the block is made of isolated unit of disks, the theorem follows immediately.  □

**Bifurcation Theorem**

If we stack any number of disks (not necessarily sequential) from Source peg to any other peg following the puzzle's rules in minimum number of moves, the number of moves required before moving the biggest disk is equal to the number of moves required after moving the biggest disk.

*Proof.* We will prove this theorem by negation. Let the number of minimum moves required before and number of moves required after, for any number of disks, are $B$ and $A$ respectively. If they are not equal, suppose $B > A$, then the strategy followed after moving the biggest disk should be taken before moving the biggest disk to keep number of moves minimum. It is possible because the situation before moving the biggest disk is same after moving the biggest disk by Blank Peg Theorem (before moving the biggest disk, the source and destination peg are usable, and it is also true after moving the biggest disk).

At the very beginning, all disks are sequential to Source peg. Before moving the biggest disk, all disks above the biggest disk are stacked to other pegs in certain arrangement. After moving the biggest disk, we can rearrange all smaller disks to the biggest disk. Clearly, the number of moves required before to transfer the sub-tower is equal to moves required after. Now, if we consider some disks to be stacked on the biggest disk with required moves $B$ and $A$, then we can follow the less moves situation accordingly. Because smaller disks above any bigger disk can be moved without any interruption to bigger disk, again, moving any bigger disk to other peg requires all smaller disks stacked in the same peg to evacuate first. So by Block-Unit theorem, $B > A$ is not possible. The proof of negation of $A > B$ is similar. $\square$

Now we are going to prove some corollaries immediately follow Bifurcation Theorem:

Let we divide first *n-1* disks sequentially upside down into $m$ blocks where $B_i^x$ denotes number of minimum moves required before moving the largest disk for *ith* block containing $x$ disks, and $A_j^y$ denotes number of minimum moves required after moving the largest disk for *jth* block containing $y$ disks . In both cases, the largest disk doesn't belong to blocks. Let *arg(.)* denotes content within a block. If $i = j$, then we assume $x = y$ and $arg(B_i^x) = arg(A_j^y)$. It ensures a particular block to be unchangeable throughout the process. For example, $arg(B_1^{n-1}) = arg(A_1^{n-1})$.

**Corollary 1:** $A_1^{n-1} = B_1^{n-1}$

*Proof:* If $x = y = n-1$, then $m = 1$. By Bifurcation Theorem, we can conclude:

$$A_1^{n-1} = B_1^{n-1}$$

**Corollary 2:** $A_1^x = B_1^x$

*Proof:* First block contains first $x$ disks. By Bifurcation Theorem, we can conclude:

$$A_1^x = B_1^x$$

**Corollary 3:** $A_1^1 = B_1^1$

*Proof:* From Corollary 2, we find for $x = 1$:

$$A_1^1 = B_1^1$$

**Corollary 4:** $B_1^{x+1} = 2B_1^x + 1$

*Proof:* By Blank Peg Theorem and Bifurcation Theorem, we can show: $B_1^{x+1} = 2B_1^x + 1$

**Corollary 5:** $B_1^2 = 2B_1^1 + 1$

*Proof:* From Corollary 4, we find for *x=1:*

$$B_1^2 = 2B_1^1 + 1$$

**Corollary 6:** (i) $B_1^m = \sum_{n=1}^m B_n^1$ (ii) $A_1^m = \sum_{n=1}^m A_n^1$

*Proof:* By Block-Unit Theorem, it follows: (i) $B_1^m = \sum_{n=1}^m B_n^1$ (ii) $A_1^m = \sum_{n=1}^m A_n^1$

**Corollary 7:** $A_x^1 = B_x^1$ where $0 < x < n$

*Proof:* By definition, it follows immediately: $A_x^1 = B_x^1$

**Reversal Theorem**

After moving whole tower, if we interchange Source peg to Destination peg and intermediates two pegs, and continue the puzzle, minimum number of moves required for both original and modified problems are same for every case if we are strict to puzzle constraints.

*Proof.* After interchanging Source peg to Destination peg and intermediates two pegs, the problem is same as original problem, and the theorem follows immediately.  □

Let $T_n$ and $\mathbf{T_n}^/$ are minimum number of moves required to transfer *n* disks from source to destination in original and modified problems respectively.

Let we divide the modified problem's first *n-1* disks sequentially into *m* blocks where $D_i^x$ denotes number of minimum moves required before moving the largest disk for *ith* block containing *x* disks, and $C_j^y$ denotes number of minimum moves required after moving the largest disk for *jth* block containing *y* disks. Let *arg(.)* denotes content within a block. If $i = j$, then we assume $x = y$ and $arg(D_i^x) = arg(C_j^y)$. It ensures a particular block to be unchangeable throughout the process. For example, $arg(D_1^{n-1}) = arg(C_1^{n-1})$.

**Corollary 8:** $A_1^x = B_1^x = C_1^x = D_1^x$

*Proof.* By Reversal Theorem, Block-Unit theorem, and Corollary 2, it follows: $A_1^x = B_1^x = C_1^x = D_1^x$

**Corollary 9:** $A_x^1 = B_x^1 = C_x^1 = D_x^1$ where $0 < x < n$

*Proof.* By definition, Block-unit Theorem, and Corollary 7, we can prove: $A_x^1 = B_x^1 = C_x^1 = D_x^1$

**Corollary 10:** For fixed *x, y,* $A_y^x = B_y^x = C_y^x = D_y^x$ where $0 < y < n$

*Proof.* From Corollary 8 and 9, it can be shown: $A_y^x = B_y^x = C_y^x = D_y^x$

**Corollary 11:** $T_{n-1} = T_{n-1}{}^{/}$

*Proof.* If we consider first *n-1* disks, according to Reversal Theorem:

$$T_{n-1} = T_{n-1}{}^{/}$$

**Definition**

For $k \in \mathbf{N}$, $k \geq 3$, let *T(n,k)* be a function which returns the minimum number of moves to solve the "Tower of Hanoi" for *n* disks on *k = 3* pegs, and the "Reve's Puzzle" for *n* disks on *k = 4* pegs respectively. We consider min{.} gives minimum number of moves on certain input-output relation.

**Recursive Formula**

According to Bifurcation theorem to solve the generalized "Tower of Hanoi," number of moves required before moving the last disk and after moving the last disk is equal, say, *S(n-1,k)*. Then the recursive formula of the generalized "Tower of Hanoi" solution for *n* disks is as follows:

$$T(0,k) = 0;$$

$$T(n,k) = 2S(n-1,k) + 1, \text{ for } n > 0, k \geq 3;$$

$$\text{If } k = 3, S(n-1,3) = T(n-1,3).$$

**Our Algorithm**

The algorithm given below was actually described by Frame. Though it is little bit complex than Stewart's one, we are working with Frame's one because it is easier to prove its optimality. The algorithm is presented below:

- Move first *l* disks using *4* pegs to *peg2* with *T(l,4)* moves.
- Move later *m* disks using *3* pegs to *peg3* with *T(m,3)* moves.
- Move last disk to *peg4* with *1* move.
- Move previous *m* disks using *3* pegs to *peg4* with another *T(m,3)* moves.
- Move remaining *l* disks using *4* pegs to *peg4* with last *T(l,4)* moves.

According to our algorithm, the recursive formula for solution to the "Reve's Puzzle" is as follows:

$$T(0,4) = 0;$$

$$T(n,4) = \min \{2T(l,4) + 2T(m,3)\} + 1, \text{ for } n > 0.$$

**Optimality of Our Algorithm**

**Lemma 1.** For *n* disks minimum number of moves required in the original " Tower of Hanoi" is greater or equal minimum number of moves required in the "Reve's Puzzle".

*Proof.* We compute successively, *T(0,3) = T(0,4) = 0; T(1,3) = T(1,4) = 1; T(2,3) = T(2,4) = 3; T(3,3) = $2^3$ - 1 = 7 ≥ 5 = T(3,4)*, and we assume: *T(n-1,3) ≥ T(n-1,4)*.

It gives:

$$T(n,3) = 2T(n-1,3) + 1 \geq 2T(n-1,4) + 1 \geq T(n,4), \text{ for } n \geq 0. \quad \square$$

**Lemma 2.** In the "Reve's Puzzle," the minimum number of moves required moving *n-1* disks to *peg2* and *peg3* must be equal to move first *l* disks using *4* pegs to *peg2* plus next *m* disks using *3* pegs to *peg3*.

*Proof.* We transfer first *l* disks using *4* pegs to *peg2* (requiring *T(l,4)* moves), and finally transfer the next *m* disks using *3* pegs to *peg3* (requiring *T(m,3)* moves). Thus we can transfer *n-1* disks in at most *T(l,4) + T(m,3)* moves, and the minimum number of moves required is *S(n-1)* :

$$S(n-1) \leq T(l,4) + T(m,3), \text{ for } n > 0.$$

But is there a better way? The answer is no. If $S_2(l)$ and $S_3(m)$ are minimum number of moves required moving *l* disks to *peg2* and *m* disks to peg3 respectively, $S(n-1) = \min \{S_2(l) + S_3(m)\}$, and the arrangement doesn't need to be sequential. But minimum number of moves required moving any *l* disks using *3* pegs to *peg2* is greater or equal moving any *l* disks using *4* pegs to *peg2*, i.e., *T(l,3) ≥ T(l,4)* by lemma 1. So we consider the definition of *T(l,4)*. It gives:

$$S_2(l) \geq T(l,4), \text{ for } l > 0.$$

Again $S_3(0) \geq 0 = T(0,3)$, and we assume: $S_3(m-1) \geq T(m-1,3)$. Now moving the *m*th disk to *peg3*, and transfer *m-1* smallest disks to *peg3* are required total of $2S_3(m-1) + 1$ moves:

$$S_3(m) = 2S_3(m-1) + 1 \geq 2T(m-1,3) + 1 = T(m,3), \text{ for } m > 0.$$

We have $S(n-1) = \min \{S_2(l) + S_3(m)\} \geq \min \{T(l,4) + T(m,3)\}$, for *n > 0*.

Finally we can conclude that:

$$S(n-1) = \min \{T(l,4) + T(m,3)\}, \text{ for } n > 0. \quad \square$$

**Strategy**

We are now in a position to define the problem formally and explore all strategies to solve the problem. Our job is to take the best strategy and prove its optimality.

Let we define four pegs as Source, Destination, Mediator, and Reservoir. As per puzzle constraints, all pegs are identical and our task is to transfer whole tower Source to Destination in minimum number of moves. As only source is fixed, we can think rest of any pegs changeably as Destination, Mediator, and Reservoir. We have generalized the problem with *n* disks. Before moving any disk at least a peg must be blank according to Blank peg theorem. Now considering Bifurcation theorem, we can proceed with following strategies:

**1st Strategy** ($S_1$): (Original Hanoi Strategy)

We take the "Reve's Puzzle" as a Tower of Hanoi problem, i.e., we totally ignore the existence of one peg, say Reservoir and always keep it untouched. As Tower of Hanoi is solvable, by Corollary 2 of Bifurcation theorem number of minimum moves required using $S_1$ is:

$$T(n,S_1) = 2T(n-1,S_1) + 1$$

**2nd Strategy** ($S_2$): (Divide and Conquer Strategy)

We take the "Reve's puzzle" as a Divide and Conquer problem, i.e., we use four pegs to stack any number of disks to a single peg before moving next disk then we transfer previous disks with equal number of moves. We can use induction method that this approach is a solution, so minimum moves required using $S_2$ is:

$$T(n,S_2) = 2T(n-1,S_2) + 1$$

We compute successively $T(0,S_1) = T(0,S_2) = 0$; $T(1,S_1) = T(1,S_2) = 1$; $T(2,S_1) = T(2,S_2) = 3$; $T(3,S_1) = T(3,S_2) = 7$; $T(4,S_1) = 15 = T(4,S_2)$ and we assume $T(n-1,S_1) = T(n-1,S_2)$. By Lemma 2, we can conclude:

$$T(n,S_1) = 2T(n-1,S_1) + 1 = 2T(n-1,S_2) + 1 = T(n,S_2)$$

So we can conclude that solving the "Reve's puzzle" as Divide and Conquer problem in minimum number of moves will always return the original "Tower of Hanoi" number, i.e., $2^n-1$.

**3rd Strategy** ($S_3$): (Dual Hanoi Strategy)

We take the "Reve's puzzle" as dual Hanoi problem, i.e., we use three pegs to stack some disks to Mediator without touching Reservoir, and then other disks to Reservoir without touching Mediator. If we want to move any disk from source and keep one peg useable for next disk, we need only one stack of disks on Mediator or Reservoir to transfer without touching other one. We continue this procedure before the biggest disk, after that we move the biggest disk to blank peg using one move and transfer all the disks upside the biggest disk with equal number of moves.

For odd numbers, $n = 2m + 1$, we can prove that minimum number of moves required using $S_3$ is:

$$T(n,S_3) = 2S(n-1,S_3) + 1 = 4T(m,3) + 1`$$

*Proof:* If $0 < l < m$, $m+l+1 \leq 2m$; hence $2^{m+l+1} \leq 2^{2m}$ which implies $2.(2^m +1) < 2^{2m-l} + 2^l + 2$. So $2T(m,3) < T(2m-l,3) + T(l,3)$. If $m < l < 2m$, $m+1 \leq 2m$; hence $2^{m+1} \leq 2^{2m-l+1}$ Which implies $2.(2^m +1) \leq 2^{2m-l} + 2^l + 2$. So $2T(m,3) \leq T(2m-l,3) + T(l,3)$.

We can conclude from above arguments that $T(n-1,S_3) = 2T(m,3)$. Hence

$$T(n,S_3) = 2S(n-1,S_3) + 1 = 4T(m,3) + 1$$

For even numbers, $n = 2m$, we can prove that minimum number of moves required using $S_3$ is:

$$T(n,S_3) = 2S(n-1,S_3) = 2(T(m,3) + T(m-1,3)) + 1$$

*Proof:* We have already prove, for $n = 2m + 1$, $T(n-1,S_3) = 2T(m,3)$. For even numbers, n = 2m, n-1 = 2m-1 = 2(m-1) + 1. So, from above formula, for n-1 = 2(m-1) + 1, we get $T(n-2,S_3) = 2T(m-1,3)$, hence $T(n-1,S_3) = 2T(m-1,3) + T(m-1,3) + 1 = T(m,3) + T(m-1,3)$. So

$$T(n,S_3) = 2S(n-1,S_3) = 2(T(m,3) + T(m-1,3)) + 1$$

Now for odd numbers, $n = 2m + 1$, $T(n,S_2) = T(2m+1,S_2) = 2^{2m+1} - 1 = 2.4^m - 1 \geq 4.2^m - 3 = 4T(m,3) + 1 = T(n,S_3)$. For even numbers, $n = 2m$, $T(n,S_2) = 2^{2m} - 1 = 4^m - 1 \geq 3(2^m - 1) = T(n,S_3)$.

**Final Strategy**: (Optimal Strategy)

Our final strategy follows Frame's algorithm partially that is we used four pegs to stack *l* disks to Mediator, then *m* disks to Reservoir, use one move for the biggest disk transferring to Destination, and finally, repeat first two steps to transfer the whole tower. Here we have totally ignored the sequence of disks. We will try to clear this issue at the last of this paper. According to optimal strategy, number of moves required:

$$T(optimal) = T(n,4) = \min \{2T(l,4) + 2T(m,3)\} + 1, \text{ for } n > 0.$$

**Proof of Optimality**

Our algorithm must be a solution of "The Reve's Puzzle". Hence

$$T(n,4) \leq 2T(l,4) + 2T(m,3) + 1, \text{ for } n > 0$$

But is there a better way? The answer is no. At some point we must move the largest disk. Then the *n-1* smallest disks must be on *peg2* and *peg3* not necessarily sequential, but they must be moved in minimum required numbers. So sequentially moving first *l* disks using *4* pegs to *peg2* and next *m* disks using *3* pegs to *peg3*, $S(n-1)$ is the minimum number required by lemma 2. Then we move the largest disk to *peg4*. Finally we need another $S(n-1)$ moves to transfer the *n-1* smallest disks back onto *peg4*:

$$T(n,4) \geq 2S(n-1) + 1 = \min \{2T(l,4) + 2T(m,3)\} + 1, \text{ for } n > 0$$

These two inequalities, together with the trivial solution for $n = 0$, yield

$$T(0,4) = 0;$$

$$T(n,4) = \min \{2T(l,4) + 2T(m,3)\} + 1, \text{ for } n > 0. \quad \square$$

**Frame's Original Definition**

For $k \geq 4$ and $n \geq k$, $T(n,k)$ is the minimum of following set:

$\{2T(n_1,k) + 2T(n_2,k-1) + ... + 2T(n_{k-2},3) + 1 \mid$

$n_1 + n_2 + ... + n_{k-2} + 1 = n, n_1 \geq n_2 \geq ... \geq n_{k-2} \geq 1\}$

where $n_i \in \mathbf{N}$

**Race to Generalized Tower of Hanoi**

As far we are only concerned with a special case of generalized "Tower of Hanoi" problem, namely four-peg "Tower of Hanoi" or the "Reve's puzzle". We have proved that our algorithm due to Frame is optimal in terms of minimum number of moves. Now we are in a position to deal with "Tower of Hanoi" problem with more than four pegs. As Frame's algorithm is proven optimal for four pegs, it indicates it might be positive for higher degrees. Basically the American Mathematical Monthly's 1939 problem was about generalized Tower of Hanoi, and both Frame and Stewart suggested solutions accordingly. Though we have encountered a special case, it can be easily generalized. Lemma 1 and 2 are the keys. If we can succeed, we will prove that "Presumed Optimal" solution indeed gives minimum number of moves.

**Uniqueness of Solution**

The original "Tower of Hanoi" of three pegs has unique solution because there is no scope to break the sequence of disks accordingly Blank Peg Theorem. It is clear that our algorithm which is sequential in nature is a solution of the "Reve's puzzle". If there are other solutions, we have to break the sequence of disks. For the "Reve's puzzle", If we consider more than one solution, we have to break the sequence of disks at some point. Let we use Mediator for four-peg moves and Reservoir for three-peg moves. Now if we break sequence before $l$ disks, we fail to achieve min{.}. So the solution of "Reve's puzzle" is sequential and unique for any disks. The possibility of unique solution of generalized "Tower of Hanoi" problem with $n$ disks and $k$ pegs is a future challenge to mathematicians with high probability of positive answer.